\documentclass[aps,prb,twocolumn,amsmath,amssymb,superscriptaddress,floatfix]{revtex4}
\usepackage{graphicx}
\usepackage{bm}
\bibstyle{apsrev.bib}

\newcommand{\be}{\begin{equation}}
\newcommand{\ee}{\end{equation}}
\newcommand{\beqn}{\begin{eqnarray}}
\newcommand{\eeqn}{\end{eqnarray}}

\begin{document}

\title{Critical behavior and entanglement of the random transverse-field Ising model\\
between one and two dimensions}

\author{Istv\'an A. Kov\'acs}
\email{ikovacs@szfki.hu}
\affiliation{Department of Physics, Lor\'and E\"otv\"os University, H-1117 Budapest,
P\'azm\'any P. s. 1/A, Hungary}
\affiliation{Research Institute for Solid State Physics and Optics,
H-1525 Budapest, P.O.Box 49, Hungary}
\author{Ferenc Igl\'oi}
\email{igloi@szfki.hu}
\affiliation{Research Institute for Solid State Physics and Optics,
H-1525 Budapest, P.O.Box 49, Hungary}
 \affiliation{Institute of Theoretical Physics,
Szeged University, H-6720 Szeged, Hungary}
\date{\today}

\begin{abstract}
We consider disordered ladders of the transverse-field Ising model and study their
critical properties and entanglement entropy for varying width, $w \le 20$, by
numerical application of the strong disorder renormalization group method. We demonstrate
that the critical properties of the ladders for any finite $w$ are controlled by the infinite disorder fixed
point of the random chain and the correction to scaling exponents contain information about the two-dimensional
model. We calculate sample dependent pseudo-critical points and study the shift of the mean values as well as scaling of the
width of the distributions and show that both are characterized by the same exponent, $\nu(2d)$.
We also study scaling of the critical magnetization, investigate critical dynamical scaling
as well as the behavior of the critical entanglement entropy. Analyzing the $w$-dependence of the results we have obtained accurate estimates for the
critical exponents of the two-dimensional model: $\nu(2d)=1.25(3)$, $x(2d)=0.996(10)$ and
$\psi(2d)=0.51(2)$.

\end{abstract}


\maketitle

\section{Introduction}
\label{sec:intro}

In nature there are materials, which are in a way between two integer dimensions, such as they are
built from $(d-1)$-dimensional layers having a finite width, $w$.
Examples are thin films\cite{thinfilms}, magnetic multilayers\cite{majkrzak91} or ladders of quantum spins\cite{dagotto96}. One interesting question for
such multilayer systems is the properties of critical fluctuations, when the linear extent of the layers, $L$,
goes to infinity. If the system is classical having thermal fluctuations, finite-size scaling
theory\cite{fisherme,barber} can be applied. One basic observation of
this theory is that for any finite $w$ the critical behavior
is controlled by the fixed point of the $(d-1)$-dimensional system, but the scaling functions
in terms of the variable, $w/L$, involve also the critical exponents of the $d$-dimensional system. For
example the critical points, $T_c(w)$, measured at a finite width, $w$, approach the true $d$-dimensional
critical point, $T_c \equiv T_c(\infty)$, as
\be
T_c-T_c(w) \sim w^{-1/\nu_{s}}\;,
\label{Tshift}
\ee
where the shift exponent, $\nu_{s}$, generally corresponds to the correlation-length exponent, $\nu$, in the $d$-dimensional system.

In a quantum system having a quantum critical point at zero temperature, $T=0$, by varying a control parameter, $\theta$, the dimensional
cross-over is a more subtle problem. If the $d$-dimensional critical quantum system is isomorphic with a $(d+1)$-dimensional
classical system\cite{kogut}, then results of finite-size scaling can be transferred to the quantum system, too. This is the case, e.g. for the
quantum critical point of the $d$-dimensional transverse-field Ising model, which is equivalent to the critical point of the classical
$(d+1)$-dimensional Ising model. However,
the situation is more complicated for antiferromagnetic models with continuous symmetry, such as for Heisenberg antiferromagnetic spin ladders.
In this case the form of low-energy excitations could sensitively depend on the value of $w$:
if the ladder contains even number of legs there is a gap, whereas for odd number of legs the system is gapless\cite{dagotto96}.
In the following for quantum systems we restrict ourselves to models with a discrete symmetry, such as to the transverse-field Ising model.

In disordered systems, in which besides deterministic (thermal or quantum) fluctuations there are also disorder fluctuations in a
sample of finite width one can define and measure a sample-dependent pseudo-critical point, $T_c(w)$ (or $\theta_c(w)$), and study its distribution\cite{domany}.
In particular one concerns the shift of the mean value, $\overline{T_c}(w)$, and the scaling of the width of the distribution, $\Delta T_c(w)$. In this case besides
the shift exponent, $\nu_{s}$, which is defined analogously to Eq.(\ref{Tshift}) one should determine the width exponent, $\nu_{w}$,
too, which is defined by the scaling relation:
\be
\Delta T_c(w) \sim w^{-1/\nu_{w}}\;.
\label{Twidth}
\ee
According to renormalization group theory\cite{aharony} the finite-size scaling behavior of random classical systems depends on the relevance or
irrelevance of the disorder\cite{harris}.
If the disorder represents an irrelevant perturbation at the pure system's fixed point, which happens if the correlation length exponent in
the pure system satisfies $\nu_{p}>2/d$, than for the disordered system we have $\nu_{s}=\nu_{p}$ and $\nu_{w}=2/d$ and the thermodynamic quantities
at the fixed point are self-averaging.
On the contrary for relevant disorder, which happens for $\nu_{p}<2/d$, there
is a new {\it conventional random fixed point} with a correlation-length exponent, $\nu \ge 2/d$\cite{ccfs}, and we have
$\nu_{s}=\nu_{w}=\nu$. In this fixed point there is a lack of self-averaging.
These predictions, which have been debated for some time\cite{psz}, were checked later
for various models\cite{domany,aharony,paz2,mai04,PS2005}.

For quantum systems quenched disorder is perfectly correlated in the (imaginary) time direction, therefore generally it has a more 
profound effect at a quantum critical point\cite{qsg}.
In some cases the critical properties of the random model are controlled by a so called infinite disorder fixed point\cite{fisher}, in which
the disorder fluctuations play a completely dominant r\v ole over quantum fluctuations. This happens, among others for the random transverse-field
Ising model, as shown by analytical results\cite{fisher} in $1d$ and numerical results\cite{pich98,motrunich00,lin00,karevski01} in $2d$.
Finite-size scaling has been tested for the
$1d$ model and a new scenario is observed\cite{ilrm}. The finite-size transition points, denoted by $\theta_c(L)$ in a system of length, $L$,
are shown to be characterized by two different exponents, $\nu_{s} < \nu_{w}$. This means, that
asymptotically $\Delta \theta_c(L)/[\overline{\theta_c}(\infty)-\overline{\theta_c}(L)] \to \infty$, which is just the opposite limit as known for irrelevant disorder.

In the present paper we go to the two-dimensional problem and study the finite-size scaling properties of ladders of
random transverse-field Ising models. For this investigations we use a numerical implementation of the so called strong disorder
renormalization group method\cite{im}. As in $2d$ this method is expected to be asymptotically exact in large scales. In the
numerical implementation of the method we have used efficient computer algorithms and in this way we could treat ladders with a large
number of sites: we went up to lengths $L=4096$ for $w=20$ legs and used $4 \times 10^4$ random samples.
Our aim with these investigations is threefold. First, we want to clarify the form of
finite-size scaling valid for this random quantum model. Second, using the appropriate form of the scaling Ansatz we want to calculate estimates
for the critical exponents of the 2d model. Previous studies\cite{pich98,motrunich00,lin00,karevski01,vojta09} in this respect have quite large error bars and we want to
increase the accuracy of the estimates considerably. Our third aim is to calculate also the entanglement entropy\cite{Amicoetal08} in the ladder
geometry and study its cross-over behavior between one\cite{refael} and two dimensions\cite{lin07,yu07}.

The structure of the rest of the paper is the following. The model and the method of the calculation is presented in Sec. \ref{sec:model}.
In Sec. \ref{sec:TC} finite-size transition points are calculated and their distribution (shift and width) is analyzed. In Sec. \ref{sec:crit}
we present calculations at the critical point about the magnetization and the dynamical scaling behavior. Results about the entanglement entropy are
presented in Sec.\ref{sec:entr}. Our paper is closed by
a discussion.

\section{Model and method}
\label{sec:model}

\subsection{Random transverse-field Ising ladder}

We consider the random transverse-field Ising model in a ladder geometry in which the sites, $i$ and $j$, are taken from a strip of the
square lattice of length, $L$, and width, $w$. We use periodic boundary conditions in both directions. The model is defined by the Hamiltonian:
\be
{\cal H} =
-\sum_{\langle ij \rangle} J_{ij}\sigma_i^x \sigma_{j}^x-\sum_{i} h_i \sigma_i^z
\label{eq:H}
\ee
in terms of the Pauli-matrices, $\sigma_i^{x,z}$. Here the first sum runs over nearest neighbor sites and
the $J_{ij}$ couplings and the $h_i$ transverse fields are independent random numbers, which are taken from the
distributions, $p(J)$ and $q(h)$, respectively.  For
concreteness we use box-like distributions: $p(J)=1$, for $0 \le J \le 1$ and $p(J)=0$,
for $J>1$; $q(h)=1/h_0$, for $0 \le h \le h_0$ and $q(h)=0$,
for $h>h_0$. We consider the system at $T=0$ and use $\theta=\ln h_0$ as the quantum control parameter.

In the thermodynamic limit, $L \to \infty$, the system in Eq.(\ref{eq:H}) displays a paramagnetic phase, for $\theta>\theta_c(w)$, and a
ferromagnetic phase, for $\theta<\theta_c(w)$. In between there is a random quantum critical point at $\theta=\theta_c(w)$ and we are going
to study its properties for various widths, $w$.

\subsection{Strong disorder renormalization group method}

The model is studied by the strong disorder renormalization group method\cite{im}, which has been introduced by Ma,
Dasgupta and Hu\cite{mdh} and later developed by D. Fisher\cite{fisher} and others. In this method the
largest local term in the Hamiltonian (either a coupling or a transverse field) is successively eliminated and at the
same time new terms are generated between remaining sites. If the largest term is a coupling, say $J_{2,3}=\Omega$ connecting sites $2$ and $3$,
($\Omega$ being the energy-scale at the given RG step), then after renormalization the two sites form a spin cluster with an effective moment
$\tilde{\mu}_{2,3}=\mu_2+\mu_3$, where in the starting situation each spin has unit moment, $\mu_i=1$. The spin cluster is put in an effective transverse
field of strength: $\tilde{h}_{2,3} \approx h_2 h_3/J_{2,3}$, which is obtained in second order perturbation calculation.
On the other hand, if the largest local term is a transverse-field, say $h_2=\Omega$, then site $2$ is eliminated and new couplings are generated
between each pairs of spins, which are nearest neighbors to $2$. If say $k$ and $l$ are nearest neighbor spins to $2$, than the new coupling connecting them
is given by: $\tilde{J}_{k,l} \approx J_{2,k} J_{2,l}/h_{2}$, also in second order perturbation calculation. If the sites $k$ and $l$ are
already connected by a coupling, $J_{k,l} \ne 0$, than for the renormalized coupling we take $max[J_{k,l},\tilde{J}_{k,l}]$.
This last step is justified if the renormalized couplings have a very broad distribution, which is indeed the case at infinite disorder
fixed points. The renormalization is repeated: at
each step one more site is eliminated and the energy scale is continuously lowered. For a finite system the renormalization is stopped at the
last site, where we keep the energy-scale, $\Omega^*$, and the total moment, $\mu^*$, as well as the structure of the clusters.

\subsection{Known exact results in the chain geometry}

The renormalization has special characters in the chain geometry, i.e. with $w=1$. In this case the topology of the system stays invariant under renormalization
and the couplings and the transverse fields are dual variables. From this follows that at the quantum critical point the couplings and the transverse fields
are decimated symmetrically, thus the critical point
is located at $\theta_c(1)=0$\cite{pfeuty}. The RG equations for the distribution function of the couplings and that of the transverse fields can be written in closed
form as an integro-differential equation, which has been solved analytically both at the critical point\cite{fisher} and in the off-critical region,
in the so-called Griffiths-phase\cite{i02}. Here we list the main results.

The energy-scale, $\Omega$, and the length-scale, $L$, are related as:
\be
\ln (\Omega_0/\Omega) \sim L^{\psi}\;,
\label{psi}
\ee
with an exponent: $\psi(1d)=1/2$. (Here $L$ can be the size of a finite system and $\Omega_0$ is a reference energy scale.)
The average spin-spin correlation function is defined as $G(r)=[\langle \sigma_i^x \sigma_{i+r}^x\rangle]_{\rm av}$, where $\langle \dots \rangle$
denotes the ground-state average and $[\dots]_{\rm av}$ stands for the averaging over quenched disorder. In the vicinity of the critical point $G(r)$
has an exponential decay:
\be
G(r) \sim \exp(-r/\xi)\;,
\label{G}
\ee
in which the correlation length, $\xi$, is divergent at the critical point as:
\be
\xi \sim |\theta-\theta_c|^{-\nu}\;,
\label{xi}
\ee
with $\nu(1d)=2$.
At the critical point the average correlations have a power-law decay:
\be
G(r) \sim r^{-2x},\quad \theta=\theta_c\;,
\label{G_crit}
\ee
with a decay exponent:
\be
x(1d)=(3-\sqrt{5})/4\;.
\label{x}
\ee
The average cluster moment, $\mu$, is related to the energy-scale, $\Omega$ as:
\be
\mu \sim [\ln (\Omega_0/\Omega)]^{\phi}\;,
\label{phi}
\ee
with $\phi(1d)=(1+\sqrt{5})/2$. The average cluster moment can be expressed also with the size as $\mu \sim L^{d_f}$, where the fractal dimension of the
cluster is expressed by the other exponents as:
\be
d_f=\phi \psi=d-x\;,
\label{d_f}
\ee
with $d=1$.

\section{Finite-size critical points}
\label{sec:TC}
\subsection{Results in the chain geometry}
In the chain geometry finite-size critical points are studied in Ref.\cite{ilrm}, in which they are located
by different methods, which all are based on the free-fermion mapping of the problem\cite{lsm}. The finite-size
critical points are shown to satisfy the micro-canonical condition:
\be
\sum_{i=1}^L \ln J_i=\sum_{i=1}^L \ln h_i\;,
\label{micro}
\ee
from which follows that the distribution of $\theta_c(L)$ is Gaussian with zero mean and with a mean deviation of
$\Delta \theta_c(L) \sim L^{-1/2}$. Consequently the width-exponent of the distribution is given by:
\be
\nu_{w}=\nu(1d)=2\;.
\label{nu_width}
\ee
On the other hand the shift-exponent is given by: 
\be
\nu_{s}=1\;,
\label{nu_shift}
\ee
although in some cases (c.f. for periodic boundary conditions) the prefactor of the scaling function can be vanishing.

\subsection{The doubling method}

In the ladder geometry, i.e. for $w \ge 2$, the free-fermionic mapping is no longer valid, therefore new methods have to be
utilized to locate pseudo-critical points. Here we used the doubling method combined with the strong disorder renormalization
group.

In the doubling procedure\cite{PS2005} for a given random sample 
($\alpha$) of length $L$ and width $w$, we construct a replicated sample ($2\alpha$) of length
$2L$ and width $w$ by gluing two copies of ($\alpha$) together and study the ratio of the
magnetizations: $r_m(\alpha,L,w)=m(2\alpha,2L,w)/m(\alpha,L,w)$, which are calculated by the strong disorder renormalization
group method. In Fig.\ref{fig1} we illustrate the $\theta$ dependence of the total magnetic moments, $\mu(2\alpha,2L,w)$ and $\mu(\alpha,L,w)$,
for a given sample of a $w=2$-leg $L=128$ ladder. The corresponding ratio of the magnetizations, $r_m(\alpha,L,w)$, is shown in the
upper inset of this figure. It is seen, that in the ordered phase: $\theta<\theta_c(\alpha,L,w)$
this ratio approaches
$r_m(\alpha,L,w) \to 1$. On the other hand in the disordered phase: $\theta>\theta_c(\alpha,L,w)$ the magnetizations approach
their minimal values, which in the SDRG method can be $1/2L$ and $1/L$, respectively, thus
we have $r_m(\alpha,L,w) \to 1/2$. In between there is a sudden change in the value of this ratio, which can be used to
define a sample-dependent pseudo-critical point, $\theta_c(\alpha,L,w)$.

There is another possibility, if we consider the ratio of the two gaps: $r_{\Omega}(\alpha,L,w)=\Omega(2\alpha,2L,w)/\Omega(\alpha,L,w)$, which
are also calculated by the strong disorder renormalization group method. In Fig.\ref{fig1} we show the two log-gaps, $-\log \Omega(2\alpha,2L,w)$ and
$-\log \Omega(\alpha,L,w)$, for the same sample as before and the corresponding ratio, $r_{\Omega}(\alpha,L,w)$, is put in the upper inset
of this figure. It is seen that this ratio in the ordered phase, $\theta<\theta_c(\alpha,L,w)$,
approaches $r_{\Omega}(\alpha,L,w) \to 0$ and in the disordered phase: $\theta>\theta_c(\alpha,L,w)$, goes to $r_{\Omega}(\alpha,L,w) \to 1$. In between
this ratio has a quick variation and we can fix the point where $r_{\Omega}(\alpha,L,w) = 1/2$ to define 
a sample-dependent pseudo-critical point, $\theta_c(\alpha,L,w)$.

\subsection{Numerical results}

\subsubsection{Comparison of the two definitions}

\begin{figure}[h!]
\begin{center}
\includegraphics[width=3.4in,angle=0]{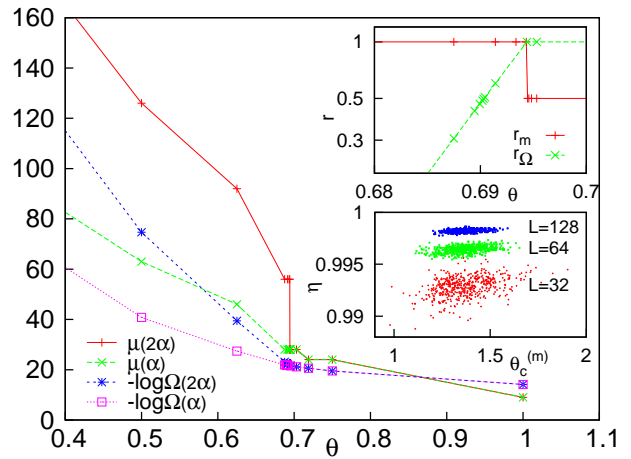}
\end{center}
\caption{
\label{fig1} (Color online)
SDRG results for the total magnetic moments, $\mu(2\alpha)$ and $\mu(\alpha)$, as well as for the log-gaps, $-\log \Omega(2\alpha)$ and $-\log \Omega(\alpha)$,
as a function of the control parameter, $\theta$, for a given realization ($\alpha$) of a $w=2$-leg $L=128$ ladder and its double ($2\alpha$), see text.
Upper inset: ratio of the magnetizations, $r_m$, and that of the gaps, $r_{\Omega}$, as a function of $\theta$ (log-lin scale) in the vicinity of the finite-size
transition points. The finite-size critical point for $r_m$ is given at the jump, for $r_{\Omega}$ it is located where $r_{\Omega}=1/2$. Lower inset:
Ratio of the two pseudo-critical points, $\eta=\theta_c^{(\Omega)}/\theta_c^{(m)}$ as a function of $\theta_c^{(m)}$ for the $w=10$ leg ladder for various
lengths and for 500 realizations.}
\end{figure}

In the doubling method we have calculated pseudo-critical points by using both ratios. We have observed, that for a given sample $\theta_c^{(\Omega)}$ calculated
from the ratio of the gaps is always somewhat smaller, than $\theta_c^{(m)}$, which is obtained from the ratio of the magnetizations. This is illustrated in the upper
inset of Fig.\ref{fig1} for a given sample. We have also calculated for several realizations
the ratio of the two pseudo-critical points, $\theta_c^{(\Omega)}/\theta_c^{(m)}$, which is shown in the lower inset of Fig.\ref{fig1} as a function of 
$\theta_c^{(m)}$ for the $w=10$ leg ladder for various
lengths, $L=32,64$ and $128$. The relative difference between the two pseudo-critical points is indeed vary small, it is of the order of $10^{-3}$ and
this is decreasing with increasing $L$ and $w$. In the following we restrict ourselves to those pseudo-critical points, which are calculated from
the ratio of the magnetization and which have a relative precision of $10^{-4}$ for each sample.


\subsubsection{Distribution of finite-size critical points}

\begin{figure}[h!]
\begin{center}
\includegraphics[width=3.4in,angle=0]{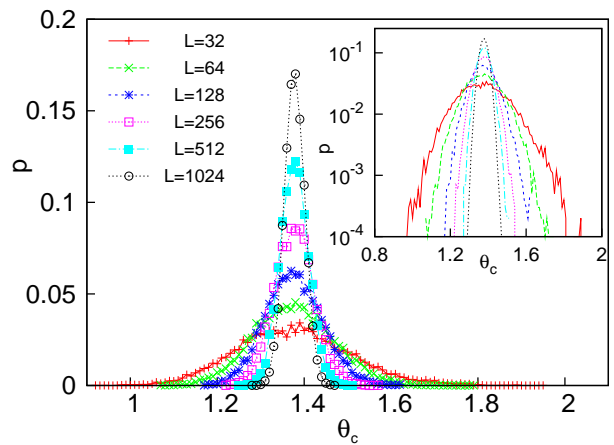}
\end{center}
\caption{
\label{fig2} (Color online)
Distribution of the pseudo-critical points, $\theta_c$, for the $w=10$ leg ladder for various
lengths and for $10^4$ realizations. In the inset in a log-lin plot deviations from the Gaussian distributions are seen,
which however are decreasing with increasing $L$.}
\end{figure}

We have calculated the distribution of pseudo-critical points for ladders with a fixed number of legs, $1 \le w \le 20$, for
varying lengths, $L=2^l$, with $l=5,6,\dots,10$. Indeed, for the largest values of $L$ the relation, $w/L \ll 1$ is well satisfied.
For the $w=10$ leg ladder the distribution of the $\theta_c$ values for various lengths are shown in Fig.\ref{fig2}, which are obtained
for $10^4$ realizations for each cases. As seen in this figure the width of the distribution is decreasing with increasing $L$ and
there is only a weak shift of the position of the maximum. The distributions somewhat deviate from Gaussians, they are asymmetric, as can be seen in the
log-lin plot in the inset of Fig.\ref{fig2}. With increasing $L$, however, the skewness of the distribution is decreasing, which is in
agreement with the expectation, that in the $w/L \to 0$ limit we get back the corresponding results for chains.


\subsubsection{"True" critical points for ladders}

For a fixed value of the number of legs, $w$, we have calculated the mean value of the pseudo-critical points. 
We have observed that the $L$-dependence of $\overline{\theta_c}(w,L)$ becomes weaker and weaker with increasing $L$, which
is in agreement with the fact, that the system approaches more and more the chain geometry. Due to this one can obtain accurate estimates
in the thermodynamic limit
for the "true" critical points of ladders, which are listed in Table \ref{table1} for different number of legs. Here the errors are merely
due to disorder fluctuations since for $L \ge 192$ the finite length effects are negligible.

\begin{table}
\caption{Quantum critical points of ladders of the random transverse-field Ising model, $\theta_c$, and the asymptotic
prefactor of the standard deviation in Eq.(\ref{Delta_scal1}), $a$,  for
different number of legs.\label{table1}}
 \begin{tabular}{|c|c|c|c|c|c|}  \hline
  $w$ & $\theta_c$ &  $a$ & $w$ & $\theta_c$ &  $a$ \\ \hline
  $1$ & $0.00021(30)$ & $1.413(8)$ & $11$ & $1.39399(20)$ & $0.728(5)$\\
  $2$ & $0.64418(15)$ & $0.997(1)$ & $12$ & $1.41211(10)$ & $0.709(3)$\\
  $3$ & $0.94736(10)$ & $0.925(5)$ &  $13$ & $1.42778(10)$ & $0.696(2)$ \\
  $4$ & $1.08059(15)$ & $0.881(2)$ &  $14$ & $1.44165(20)$ & $0.681(4)$ \\  
  $5$ & $1.16859(10)$ & $0.844(5)$ &  $15$ & $1.45397(10)$ & $0.669(4)$ \\  
  $6$ & $1.23207(15)$ & $0.815(5)$ &  $16$ & $1.46472(15)$ & $0.658(4)$ \\  
  $7$ & $1.27962(15)$ & $0.807(4)$ &  $17$ & $1.47445(10)$ & $0.650(4)$ \\  
  $8$ & $1.31727(20)$ & $0.781(3)$ &  $18$ & $1.48332(10)$ & $0.640(1)$ \\  
  $9$ & $1.34787(10)$ & $0.763(3)$ &  $19$ & $1.49095(30)$ & $0.626(4)$ \\  
  $10$& $1.37270(15)$ & $0.733(2)$ &  $20$ & $1.49855(15)$ & $0.620(2)$ \\    \hline
  \end{tabular}
  \end{table}

Here we also list our estimate for the chain, $w=1$, which agrees within the error of the calculation with the exact result: $\theta_c(1)=0$
and $a(1)=\sqrt{2}$ (see Sec.\ref{sec:width}).

\begin{figure}[h!]
\begin{center}
\includegraphics[width=3.4in,angle=0]{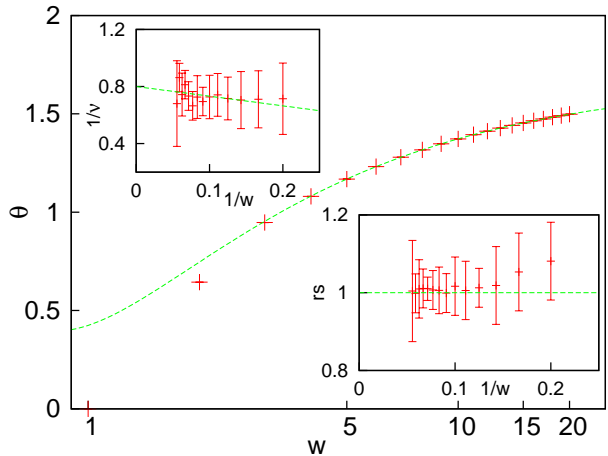}
\end{center}
\caption{
\label{fig3} (Color online)
Critical points of the ladders for varying number of legs, $w$, and the extrapolation curve (broken (green) line) for large $w$. Upper inset: estimates of the inverse of the
local shift exponent, $1/\nu_s(w)$, calculated through Eq.(\ref{nu_shift_w}). The broken (green) straight line indicates the extrapolation through $1/w$.
Lower inset: ratio of the scaled difference
of the critical points and the scaled standard deviations as a function of $1/w$(see text). The horizontal broken (green) line at $rs=1.$ indicates a value
which is close to the expected asymptotic behavior.
}
\end{figure}


These data approach the critical point in the 2d system, $\theta_c(2d)$, see Fig.\ref{fig3}. Here the corrections for large-$w$ are expected to have a power-law form,
and analogously to Eq.(\ref{Tshift}), it contains the shift exponent, $\nu_{s}$, of the 2d system.

Estimates for the effective ($w$-dependent) values of the shift exponent are obtained from the ratio of the second and the first finite differences:
\be
\frac{1}{\nu_{s}(w)}=\frac{\Delta_2\theta_c(w)}{\Delta_1\theta_c(w)}w-1\;,
\label{nu_shift_w}
\ee
which are calculated at the central point of five-point fits. The effective exponents are given in the upper inset of Fig.\ref{fig3},
which are extrapolated as $1/\nu_s(2d)=0.81(10)$, thus we obtain:
\be
\nu_s(2d)=1.24(15)\;.
\ee

\subsubsection{Scaling of the width of the distribution}
\label{sec:width}
\begin{figure}[h!]
\begin{center}
\includegraphics[width=3.4in,angle=0]{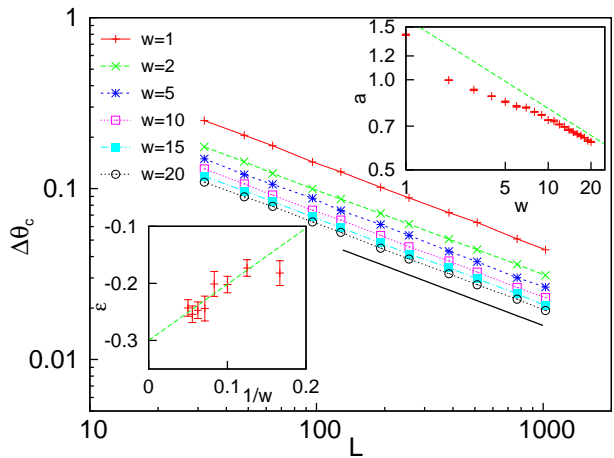}
\end{center}
\caption{
\label{fig4} (Color online)
Scaling of the width of the distribution of the pseudo-critical points, $\Delta \theta_c(w,L)$, with $L$ for different number of legs, $w$.
In a log-log plot the slope of the curves are compatible with the theoretical prediction, $\nu_w(1d)=1/2$, which is indicated by a full straight line.
In the upper inset the limiting value of
the prefactor, $a(w)$ is shown as a function of $w$ in a log-log plot. The dashed (green) straight line has a slope, $\epsilon=0.30$,
as extrapolated from effective exponents in the lower inset.}
\end{figure}

We have measured the standard deviation of the distribution of the pseudo-critical points, $\Delta \theta_c(w,L)$, for ladders with $w$ legs and
with a varying length, $L$. This quantity is expected to scale with the length as:
\be
\Delta \theta_c(w,L)=L^{-1/\nu_{w}(2d)} \sigma(w/L)\;,
\label{Delta_scal}
\ee
where the scaling function, $\sigma(y)$, behaves for small arguments as: $\sigma(y) \sim y^{-1/\nu_{w}(2d)+1/\nu_{w}(1d)}$. From this follows, that
for finite widths we have:
\be
\Delta \theta_c(w,L) = L^{-1/\nu_{w}(1d)} a(w)\;,
\label{Delta_scal1}
\ee
with a prefactor, $a(w)$, which behaves for large $w$ as $a(w)\sim w^{\epsilon}$, with an exponent $\epsilon= -1/\nu_{w}(2d)+1/\nu_{w}(1d)$. 
We have checked this scenario by analyzing the data
for $\Delta \theta_c(w,L)$. First, for a fixed $w$ we have fitted a function $a(w) L^{-\omega}$, with a free parameter, $\omega$. We have found that for
each widths, $1 \le w \le 20$, the exponent $\omega$ agrees with $1/\nu_w(1d)=0.5$, within a few thousands of error, as illustrated in Fig.\ref{fig4}.

In the next step we have fixed the value of $\omega=0.5$ and estimated the limiting value of $\Delta \theta_c(w,L)L^{0.5}$ for large $L$, which is
denoted by $a(w)$. These limiting values are presented in Table \ref{table1}, which are analyzed for large $w$. As seen in the upper inset of
Fig. \ref{fig4} in a log-log
plot the $a(w)$ values are asymptotically on a straight line. We have calculated effective, $w$-dependent exponents:
$\epsilon(w)=\log (a(w)/a(w/2))/\log 2$, which are presented in the lower inset of Fig.\ref{fig4} as a function of $1/w$. These effective exponents
have a weak $w$-dependence and we estimate its limiting value as $\epsilon=-0.30(2)$. With this we have for the width exponent in $2d$:
\be
\nu_w(2d)=1.25(3)\;.
\ee
Closing this section we try to decide in a direct way about the relation between the two exponents, $\nu_s(2d)$ and $\nu_w(2d)$. For
this we form the scaled difference: $d\theta_c(w)=\Delta_1\theta_c(w) w $ (see Eq.(\ref{nu_shift_w})), which scales as $w^{-1/\nu_s}$, as well as
the scaled standard deviation: $sa(w)=w^{-0.5} a(w)$, which scales as $w^{-1/\nu_w}$, and form their ratio, $rs(w)=d\theta_c(w)/sa(w)$.
As seen in the lower inset of Fig.\ref{fig3} this ratio approaches a finite value which can be estimated as $rs=1.01(2)$.
Thus we can conclude that at the infinite disorder fixed point of the $2d$ random transverse-field Ising model the
shift and the width exponents are equal and they correspond to the correlation length exponent of the model.

Using the best estimate for $\nu_s(2d)=\nu_w(2d)$ and including the first analytic correction to scaling term: $\theta_c=\theta_c(w)-A w^{-1/\nu_s}(1+B/w)$
we fit our data (see Fig.\ref{fig3}) and obtain for the critical point of the $2d$ system:
\be
\theta_c(2d)=1.676(5)\;.
\ee
This value is in agreement with the previous estimate, $\theta_c(2d)=1.680(5)$, in Ref.[\onlinecite{yu07}].

\section{Scaling at the critical point}
Having estimates for the critical points of random ladders with $w$ legs, $\theta_c(w)$, we have calculated scaling of the magnetization
at the critical point as well as the critical dynamical scaling. These calculations are made for lengths up to $2^{12}$ and for $4 \times 10^4$
realizations.
\label{sec:crit}
\subsection{Magnetization}

We have calculated the
average total magnetic moment at the critical point, $\mu_c(w,L)$, for varying lengths, $L$, which is expected to scale as:
\be
\mu_c(w,L)=L^{d_f(2d)} \tilde{\mu}_c(w/L)\;,
\label{mu_scal}
\ee
with a scaling function, which behaves for small arguments as: $\tilde{\mu}_c(y) \sim y^{d_f(2d)-d_f(1d)}$. Then, for a finite width, $w$,
we have:
\be
\mu_c(w,L)=L^{d_f(1d)} b(w)\;,
\label{mu_scal1}
\ee
with a prefactor, which for large $w$ behaves as: $b(w) \sim w^{\kappa}$, with $\kappa=d_f(2d)-d_f(1d)$.

\begin{figure}[h!]
\begin{center}
\includegraphics[width=3.4in,angle=0]{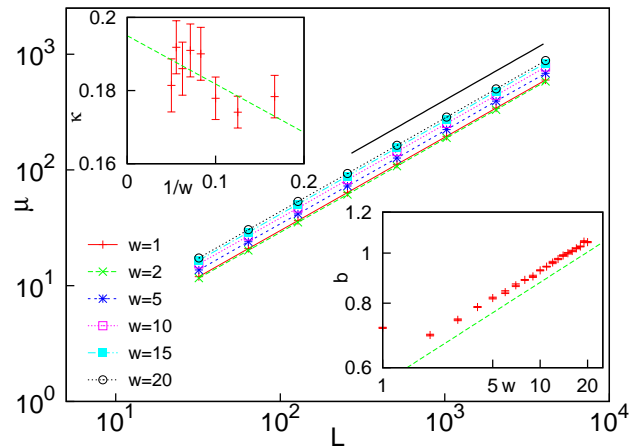}
\end{center}
\caption{
\label{fig5} (Color online)
Scaling of the average total magnetic moment at the critical point of a ladder with $w$ legs and length $L$, $\mu_c(w,L)$.
In the log-log plot the slope of the curves are compatible with the theoretical prediction, $d_f(1d)$ in Eq.(\ref{d_f}), which is indicated by a full straight line.
In the lower inset the limiting value of
the prefactor, $b(w)$ is shown as a function of $w$ in a log-log plot. The dashed (green) straight line has a slope, $\kappa=0.195$,
as extrapolated from effective exponents in the upper inset.}
\end{figure}

The scaling Ansatz in Eq.(\ref{mu_scal1})
is checked in Fig.\ref{fig5}. Then, we have calculated the limiting value of $L^{-d_f(1d)} \mu_c(w,L)$, which is denoted by $b(w)$ and which is
presented as a function of $w$ in a log-log plot in the lower inset of Fig.\ref{fig5}. Effective, $w$-dependent exponents are calculated, which are
extrapolated in the upper inset of Fig.\ref{fig5} giving $\kappa=0.195(10)$. Thus the fractal dimension in $2d$ is $d_f(2d)=d_f(1d)+\kappa$ and from Eq.(\ref{d_f})
we obtain for the magnetization scaling dimension:
\be
x(2d)=0.996(10)\;.
\ee

\subsection{Dynamical scaling}

\begin{figure}[h!]
\begin{center}
\includegraphics[width=3.4in,angle=0]{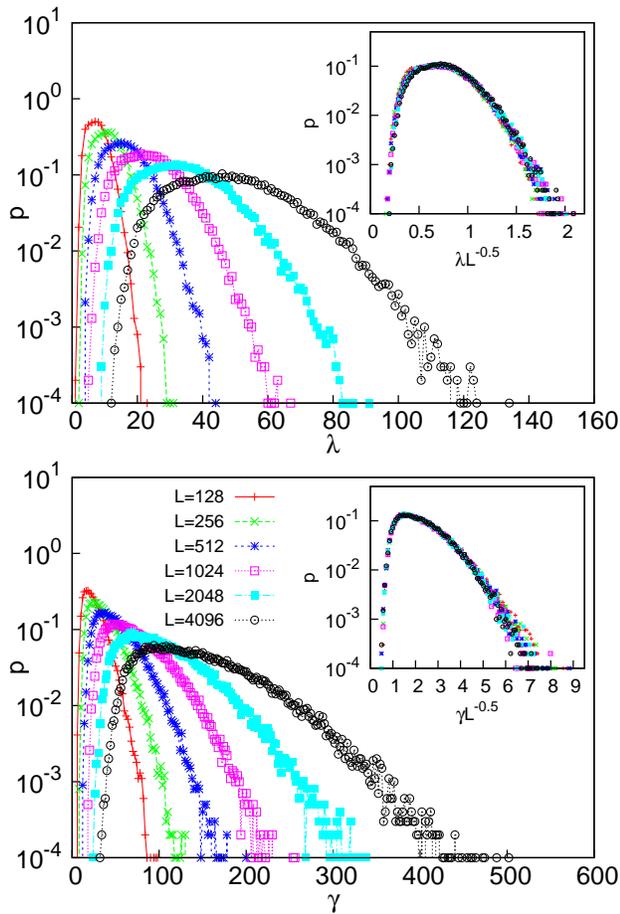}
\end{center}
\caption{
\label{fig6} (Color online)
Distribution of the last decimated log-coupling, $\lambda$, (upper panel) and the last decimated log-transverse field, $\gamma$ (lower panel) for various lengths, $L$,
for the $w=10$-leg ladder. In the insets the distribution of the scaled variables: $\lambda L^{-0.5}$ and $\gamma L^{-0.5}$, respectively are
shown.}
\end{figure}

At an infinite disorder fixed point there is a special form of dynamical scaling, as given in Eq.(\ref{psi}). The
energy scale of a sample at the end of the renormalization can be defined either by the value of the last decimated
(log) coupling $-\log \tilde{J}=\lambda$ or by the last decimated (log) transverse field $-\log \tilde{h}=\gamma$. The
distribution of $\lambda$ as well as $\gamma$ are shown in Fig.\ref{fig6} in upper and in the lower panel, respectively, for the $w=10$-leg ladder
for various values of the length, $L$. An appropriate scaling collapse of the date is observed in terms of the scaling
variables, $\lambda L^{-\psi}$ and $\gamma L^{-\psi}$, with $\psi=\psi(1d)=1/2$, as illustrated in the insets.

In order to have a more quantitative picture about dynamical scaling we consider the mean value: $\Gamma(w,L)=[\gamma(w,L)]_{\rm av}$
and the standard deviation, $\Delta \Gamma(w,L)$ and similarly, $\Lambda(w,L)=[\lambda(w,L)]_{\rm av}$ and $\Delta \Lambda(w,L)$. All these
quantities are expected to scale in the same way, for example with $\Gamma(w,L)$ we have:
\be
\Gamma(w,L)=L^{\psi(2d)} \tilde{\Gamma}(w/L)\;,
\label{Gamma_scal}
\ee
with $\tilde{\Gamma}(y) \sim y^{\psi(2d)-\psi(1d)}$. For a finite width, $w$, we have then:
\be
\Gamma(w,L)=L^{\psi(1d)} g(w)\;,
\label{Gamma_scal1}
\ee
with $g(w) \sim w^\delta$ for large $w$ with $\delta=\psi(2d)-\psi(1d)$.

\begin{figure}[h!]
\begin{center}
\includegraphics[width=3.4in,angle=0]{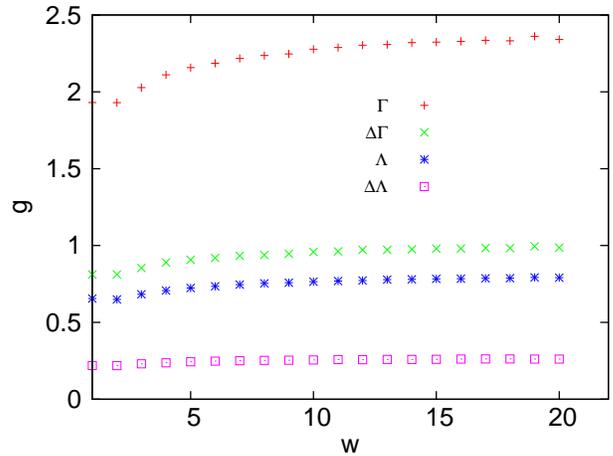}
\end{center}
\caption{
\label{fig7} (Color online)
Scaling functions for typical energy-scales: $g(w)= \lim_{L \to \infty} G(w,L)L^{-1/2}$, in which $G(w,L)$ is either $\Gamma(w,L)$, $\Delta \Gamma(w,L)$, $\Lambda(w,L)$ or
$\Delta \Lambda(w,L)$, see text.}
\end{figure}

We have checked that the scaling form in Eq.(\ref{Gamma_scal1}) is indeed
satisfied for all values of $1 \le w \le 20$ and than calculated the limiting value of $\Gamma(w,L) L^{-1/2}$, which is denoted by $g(w)$. As illustrated
in Fig.\ref{fig7} the scaling functions of the typical energy-scales have only a very weak $w$ dependence, and we estimate (not shown) a small exponent: $\delta=0.01(2)$.
Thus we have for the $\psi$ exponent in the $2d$ model:
\be
\psi(2d)=0.51(2)\;.
\ee

\section{Entanglement entropy}
\label{sec:entr}
In the ladder geometry we consider a block, $\cal{A}$, which contains all the $w$ legs and has a length, $\ell \ll L$. Consequently the block has two
parallel lines of width, $w$, at which it has contact with the rest of the system, $\cal{B}$. The entanglement of $\cal{A}$
with $\cal{B}$ is quantified by the von-Neumann entropy\cite{Amicoetal08}:
\be
{\cal S}_{\cal A}(w,\ell)=-{\rm Tr}_{\cal A}(\rho_{\cal A} \log \rho_{\cal A})\;,
\label{S}
\ee
in terms of the reduced density matrix: $\rho_{\cal A}=Tr_{{\cal B}} | \Psi \rangle \langle \Psi |$, where $| \Psi \rangle$ denotes a pure state (in
our case the ground state) of the complete system.

\begin{widetext}


\begin{figure}[h!]
\begin{center}
\includegraphics[width=6.8in,angle=0]{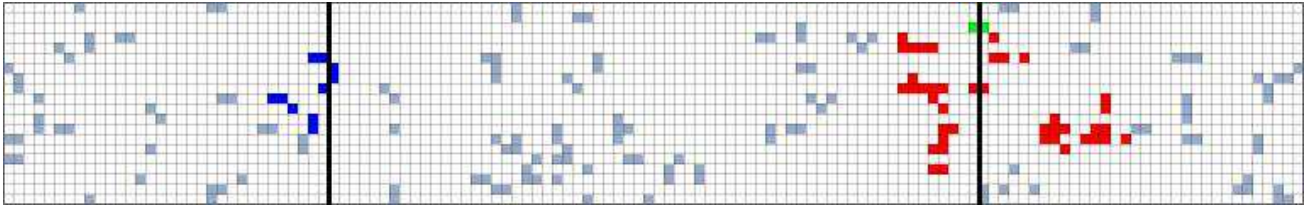}
\end{center}
\caption{
\label{fig8} (Color online)
Structure of the decimated spin clusters at the critical point of a ladder of $20 \times 128$ sites, which is devided into two equal blocks the boundary of which
is indicated by thick vertical lines. There are three clusters, denoted by blue, red and green colors, respectively, which contain sites at both blocks and thus result in
an entropy $3 \times \log 2$.}
\end{figure}


\end{widetext}

At the critical point of a random quantum system the properties of which are controlled by an infinite disorder fixed point the asymptotic behavior of
the entropy in the large $l$ limit can be obtained by the strong disorder renormalization group method. For the random transverse-field Ising model entanglement
between $\cal{A}$ and $\cal{B}$ are given by such renormalized spin clusters, which contain sites both in $\cal{A}$ and in $\cal{B}$, and the cluster is
eliminated at some point of the renormalization\cite{refael,lin07,yu07}. Due to the very broad distribution of the effective couplings and transverse fields, the cluster at the
energy scale of its decimation is in a so called GHZ entangled state of the form: $1/\sqrt{2}(|\uparrow \uparrow  \dots  \uparrow\rangle +
|\downarrow \downarrow \dots  \downarrow\rangle )$. Each such cluster contribute by an amount of $\log 2$ to the entanglement entropy, thus calculation
of the entropy is equivalent to a cluster counting problem, which is illustrated in Fig.\ref{fig8}.
In the chain geometry the asymptotic behavior of the entropy at the critical point is obtained from the analytical solution of the RG equations as\cite{refael}:
\be
{\cal S}_{\cal A}(1,\ell) \approx \frac{c(1)}{3} \log \ell + k(1)\;,
\label{S_d1}
\ee
where $k(1)$ is a non-universal constant, which depends on the form of the disorder, whereas the prefactor of the logarithm, $c(1)$, which is also
called as the effective central charge, is universal and given by: $c(1)=\ln 2/2$. This result is checked numerically in Ref.[\onlinecite{il08}].
In the two-dimensional case, which is expected to
hold for $w/L=O(1)$, there are somewhat
conflicting numerical results at the critical point.
Lin {\it et al.}\cite{lin07} have observed a double-logarithmic multiplicative factor to the area-law: ${\cal S}_{\cal A}(\ell,\ell) \approx \ell \log \log \ell$
whereas later Yu {\it et al.}\cite{yu07} argued to have only a subleading logarithmic term to the area law: ${\cal S}_{\cal A}(\ell,\ell) \approx a \ell +b \log \ell + k$.

\begin{figure}[h!]
\begin{center}
\includegraphics[width=3.4in,angle=0]{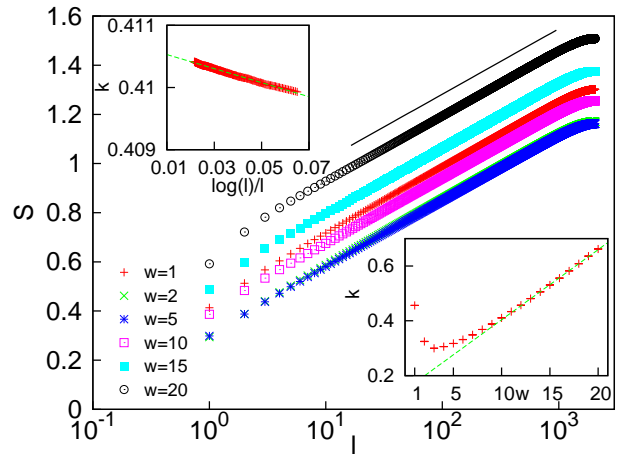}
\end{center}
\caption{
\label{fig9} (Color online)
The log-$\ell$-dependence of the entanglement entropy at the critical point of ladders for different number of legs, $w$, and for $L=4096$. 
The linear part of the curves has approximately the same slope, which is consistent with $\ln 2/6$, as indicated by a full straight line.
Upper inset: the non-universal part of the critical entropy, $k(w,\ell)$, for the $w=10$-leg ladder and its extrapolation for $\ell \gg w$
(but $\ell \ll L=4096$) with a correction term of $\sim \log\ell/\ell$. Lower inset: the asymptotic value of the non-universal part of the
critical entropy, $k(w)$, as a function of $w$. Asymptotically there is a linear $w$-dependence, which is shown by a broken (green) straight
line.}
\end{figure}

Here we study numerically the critical ladder systems with various number of legs and try to identify the cross-over between one- and two dimensions.
To illustrate the $\ell$ dependence of the entanglement entropy in Fig.\ref{fig9} we show ${\cal S}_{\cal A}$ as a function
of $\log \ell$ for different number of legs for $L=4096$. (We have checked, that the asymptotic results does not change for $L=2048$.)
The central parts of the curves are very well linear having approximately the same slope, which is consistent with the exact result
for the $w=1$ chain geometry. Thus we conclude that the effective central charge, $c(w)$, does not depend on the number of legs.

In the next step we fix $c(w)=\ln 2/2$, calculate the non-universal term: $k(w,\ell)={\cal S}_{\cal A}(w,\ell) - \frac{\ln 2}{6} \log \ell$
and take its limit, $k(w)$, for large $\ell$ (but still with $\ell \ll L$). As illustrated in the upper inset of Fig.\ref{fig9}
the $\ell$-dependent correction term is approximated as $\log \ell/\ell$ and the asymptotic non-universal terms, $k(w)$, are shown
for different number of legs in the lower inset of Fig.\ref{fig9}. One can see that starting
with the chain, $w=1$, first $k(w)$ is decreasing, has a minimum around $w=3$ and then starts to increase. This increase for large $w$ is approximately linear,
we have fitted: $k(w)=0.0256(5)w+0.148(5)$. This linear increase is compatible with the area law, which should hold for non-critical systems and for large blocks. Our
analysis can be used to clarify the one- to two-dimensional cross-over of the entropy in the limit $w/\ell \ll 1$. However, our data can not be used to make
predictions further, for $w/L=O(1)$, i.e. for the two-dimensional case. For this one should analyze the occurrent and possibly very weak $w$
dependence of the prefactor of the linear term of $k(w)$, which however can not be done with our data, which are only up to $w=20$.

\section{Discussion}
\label{sec:disc}

In this paper we have studied the critical properties and the entanglement entropy of random transverse-field Ising
models in the ladder geometry by the strong disorder renormalization group method. In our
numerical calculation we went up to w=20 legs and  with a length up to L=4096 for $4 \times 10^4$ realizations.
In principle the sizes of the systems could have been increased further, but it was not necessary. With $L$
we have already reached the limit where no further systematic finite-size effects are seen. On the other hand
for larger values of $w$ we would have obtained too large errors in calculating quantities, such as through two-point fit.

First, we have calculated sample dependent finite-size critical points, which are obtained by the doubling procedure
and the strong disorder renormalization group method. We have analyzed the shift of the mean value of the transition points and the width of the
distribution as a function of the number of legs, $w$, and estimated the exponents of the $2d$ model, $\nu_s(2d)$ and $\nu_w(2d)$, respectively.
These are found to be identical and given by the correlation-length
exponent of the $2d$ model. Consequently the scaling behavior of the pseudo-critical points of the $2d$ random
transverse-field Ising model is in the same form as for classical and conventional random fixed points\cite{aharony}. In this respect
there is a difference with the $1d$ model\cite{ilrm}, in which $\nu_s(1d)<\nu_w(1d)$. For this latter model probably the free-fermionic
character could be the reason for the different scaling properties.
Our estimate for the correlation length exponent, $\nu(2d)=1.25(3)$, is clearly larger than the possible limiting value of $2/d$,
which has been observed in the $1d$ model and in some other random systems\cite{psz}.

Scaling at the critical point for different quantities is analyzed in a similar way, what we summarize here as follows. Let
us consider a physical observable, ${\bf A}$, which at the critical point has the mean value, $A(w,L)$. This quantity
scales with the critical exponent of the $2d$ model, $\alpha(2d)$, as:
\be
A(w,L)\sim L^{\alpha(2d)} \tilde{A}(w/L)
\ee
where the scaling function, $\tilde{A}(y)$, for small arguments behaves as:
\be
\tilde{A}(y)\sim y^{\alpha(2d)-\alpha(1d)}
\ee
where $\alpha(1d)$ is the critical exponent in the $1d$ model. Consequently 
for a finite $w$, but for $L \to \infty$, we have
\be
A(w,L)\sim L^{\alpha(1d)} a(w)
\ee
with $a(w)\sim w^\omega$ and $\omega={\alpha(2d)-\alpha(1d)}$. In general we measure the scaling function $a(w)$ for
different widths, estimate the exponent $\omega$ and calculate the critical exponent in $2d$ as: $\alpha(2d)=\alpha(1d)+\omega$.
Since the exponents in $1d$ are exactly known and the correction term, $\omega$, is comparatively small we have obtained quite
accurate exponents in $2d$. In the following we compare the estimates for the different critical exponents in the $2d$ infinite
disorder fixed point, which are listed in Table \ref{table2}.

\begin{table}
\caption{Numerical estimates of the critical exponents at the infinite disorder fixed point in $2d$. MC: Monte Carlo simulation;
SDRG: numerical strong disorder renormalization group; CP: Monte Carlo simulation of the $2d$ random contact process. The
exponents, $\phi$, denoted by an asterisk are calculated from the scaling relation in Eq.(\ref{d_f}). \label{table2}}
 \begin{tabular}{|c|c|c|c|c|}  \hline
$\psi$&$\phi$&$\nu$&$x$&method\\ \hline
0.4(1)&2.5*&&1.0&MC\cite{pich98} \\
0.42(6)&2.5(4)&1.07(15)&1.0(1)& SDRG\cite{motrunich00}\\
0.5&2&&0.94&SDRG\cite{lin00}\\
0.6&1.7&1.25&0.97&SDRG\cite{karevski01}\\
0.51(6)&2.04(28)*&1.20(15)&0.96(2)&CP\cite{vojta09} \\ \hline
0.51(2)&1.97(10)*&1.25(3)&0.996(10)& this work \\   \hline
  \end{tabular}
  \end{table}

Here besides different numerical strong disorder renormalization group results there are also Monte Carlo simulations, both for the
random transverse-field Ising model and for the random contact process. This latter model is expected to belong to the same universality
class\cite{hiv}, at least for strong enough disorder. It is seen in Table \ref{table2} that our estimates fit to the trend of the previous
results and generally have a somewhat smaller error.

We have also studied the scaling behavior of the entanglement entropy in the ladder geometry. For a fixed width, $w$, the entropy
is found to grow logarithmically with the length of the block, $\ell$, and the prefactor is found independent of $w$. On the other hand the
$\ell$ independent term of the entropy is found to have a linear $w$ dependence, at least for large enough $w$, which corresponds
to the are law for this systems.

The investigations presented in this work can be naturally continued for larger and larger widths and approaching the case, $w/L=O(1)$, which
corresponds to the two-dimensional model. However, with increasing $w$ the numerical computation becomes more and more costly. The reason for this
is the fact that the connected clusters in the strong disorder renormalization group method are typically of size $w \times w$, which for large $w$ becomes
fully connected after decimating a small percent of the transverse fields. The number of further renormalization steps grows in a na\"{\i}ve approach as
$w^6$, so that by this method one can not go further than $L \sim 100$ or $200$ in $2d$. To treat larger $2d$ systems improved algorithms are
necessary. Studies in this direction are in progress.

\begin{acknowledgments}
This work has been
  supported by the Hungarian National Research Fund under grant No OTKA
K62588, K75324 and K77629 and by a German-Hungarian exchange program (DFG-MTA).
We are grateful to P. Sz\'epfalusy, H. Rieger and Y-C. Lin for useful discussions.
\end{acknowledgments}

\end{document}